\documentclass[nonacm]{acmart}

\usepackage{algorithmic}
\usepackage{braket}
\usepackage{graphicx}
\usepackage{textcomp}
\usepackage{comment}

\newcommand{\todo}{\textcolor{blue}{TODO: }\textcolor{blue}}

\usepackage{soul}
\usepackage{booktabs}
\usepackage{adjustbox}

\usepackage[linewidth=1pt]{mdframed}

\usepackage{tikz}
\newcommand*\circled[1]{\tikz[baseline=(char.base)]{
            \node[shape=circle,draw,inner sep=0.5pt] (char) {#1};}}
\usepackage{mathtools}

\usepackage[font=bf,skip=0pt]{caption}
\usepackage{tabularx}

\usepackage[caption=false]{subfig}

\AtBeginDocument{%
  \providecommand\BibTeX{{%
    \normalfont B\kern-0.5em{\scshape i\kern-0.25em b}\kern-0.8em\TeX}}}

\begin{document}

\title{Error Mitigation in Quantum Computers through Instruction Scheduling}

\author{Kaitlin N. Smith}
\email{kns@uchicago.edu}
\affiliation{%
  \institution{University of Chicago}
  \city{Chicago}
  \state{Illinois}
  \country{USA}
}

\author{Gokul Subramanian Ravi}
\affiliation{%
  \institution{University of Chicago}
  \city{Chicago}
  \state{Illinois}
  \country{USA}
}

\author{Prakash Murali}
\affiliation{%
  \institution{Princeton University}
  \city{Princeton}
  \state{New Jersey}
  \country{USA}
}

\author{Jonathan M. Baker}
\affiliation{%
  \institution{University of Chicago}
  \city{Chicago}
  \state{Illinois}
  \country{USA}
}

\author{Nathan Earnest}
\affiliation{%
  \institution{IBM Quantum, IBM T.\ J.\ Watson Research Center}
  \city{Yorktown Heights}
  \state{New York}
  \country{USA}
}

\author{Ali Javadi-Abhari}
\affiliation{%
  \institution{IBM Quantum, IBM T.\ J.\ Watson Research Center}
  \city{Yorktown Heights}
  \state{New York}
  \country{USA}
}

\author{Frederic T. Chong}
\affiliation{%
  \institution{University of Chicago}
  \city{Chicago}
  \state{Illinois}
  \country{USA}
}

\renewcommand{\shortauthors}{Smith, et al.}

\begin{abstract}
Quantum systems have potential to demonstrate significant computational advantage, but 
current quantum devices suffer from the rapid accumulation of error that prevents the storage of quantum information over extended periods. The unintentional coupling of qubits to their environment and each other adds significant noise to computation, and improved methods to combat decoherence are required to boost the performance of quantum algorithms on real machines. While many existing techniques for mitigating error rely on adding extra gates to the circuit~\cite{viola1999dynamical,giurgica2020digital,das2021adapt}, calibrating new gates~\cite{temme2017error}, or extending a circuit's runtime~\cite{murali2020software}, this paper's primary contribution leverages the gates 
already present in a quantum program without extending circuit duration. We exploit circuit slack for single-qubit gates that occur in idle windows, scheduling the gates such that their timing can counteract some errors.

Spin-echo corrections that mitigate decoherence on idling qubits 
act as inspiration for this work. Theoretical models, however, fail to capture all sources of noise in NISQ devices, making practical solutions necessary that better minimize the impact of unpredictable errors in quantum machines. 
This paper presents TimeStitch: a novel framework that pinpoints the optimum execution schedules for single-qubit gates within quantum circuits. TimeStitch, implemented as a compilation pass, leverages the reversible nature of quantum computation to boost the success of circuits on real quantum machines. Unlike past approaches that apply reversibility properties to improve quantum circuit execution~\cite{Patel2021QraftRY}, TimeStitch amplifies fidelity without violating critical path frontiers in either the slack tuning procedures or the final rescheduled circuit. On average, compared to a state-of-the-art baseline, a practically constrained TimeStitch achieves a mean 38\% relative improvement in success rates, with a maximum of 106\%, while observing bounds on circuit depth. When unconstrained by depth criteria, TimeStitch produces a mean relative fidelity increase of 50\% with a maximum of 256\%. Finally, when TimeStitch intelligently leverages periodic dynamical decoupling within its scheduling framework, a mean 64\% improvement is observed over the baseline, relatively outperforming standalone dynamical decoupling by 19\%, with a maximum of 287\%.
\end{abstract}

\maketitle

\begin{acks}
This work is funded in part by EPiQC, an NSF Expedition in Computing, under grants CCF-1730082/1730449; in part by STAQ under grant NSF Phy-1818914; in part by DOE grants DE-SC0020289 and DE-SC0020331; in part by NSF OMA-2016136 and the Q-NEXT DOE NQI Center; and in part by in part by NSF grant OMA-1936118. KNS is supported by IBM as a Postdoctoral Scholar at the University of Chicago and the Chicago Quantum Exchange. GSR is supported as a Computing Innovation Fellow at the University of Chicago. This material is based upon work supported by the National Science Foundation under grant \# 2030859 to the Computing Research Association for the CIFellows Project.  PM is supported by an IBM PhD Fellowship at Princeton University. Frederic T. Chong is Chief Scientist at Super.tech and an advisor to Quantum Circuits, Inc.

\end{acks}

\section{Introduction}
\label{1-introduction}

Quantum computing is a revolutionary computational model that leverages quantum mechanical phenomena for solving intractable problems. Quantum computers (QCs) evaluate quantum circuits, or programs, in a manner similar to classical computers, but quantum information's ability to leverage superposition, interference, and entanglement is projected to provide QCs significant advantage in cryptography~\cite{shor1999polynomial}, chemistry~\cite{kandala2017hardware}, optimization~\cite{moll2018quantum}, and machine learning~\cite{biamonte2017quantum} applications.

\begin{figure}[t!]

\begin{adjustbox}{varwidth=0.6\columnwidth,fbox,center}

\subfloat[Single-qubit gates are rescheduled for their optimum placement within slack, thereby mitigating decoherence errors.]{%
\includegraphics[width=\columnwidth,trim={0cm 11.2cm 0cm 0cm},clip]{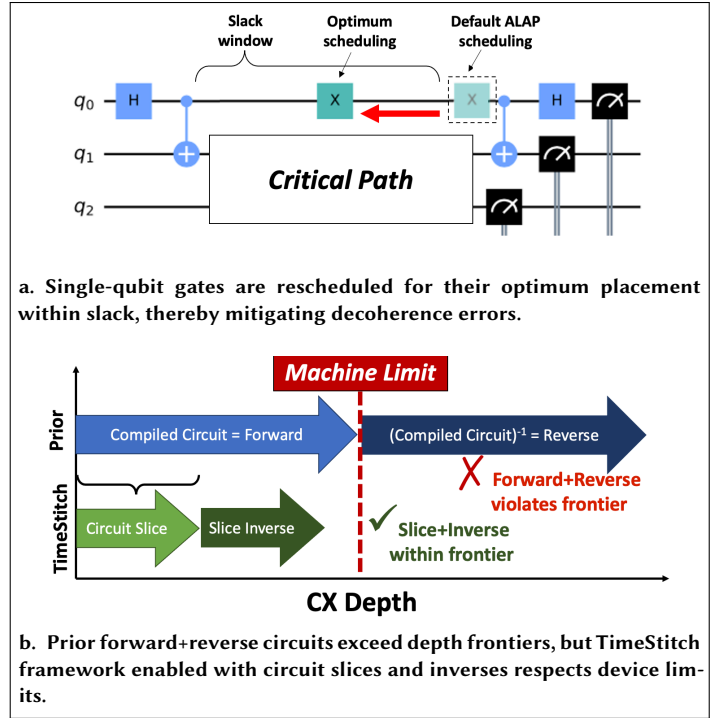}}

\subfloat[Prior forward+reverse circuits exceed depth frontiers, but TimeStitch framework enabled with circuit slices and inverses respects device limits.]{%
\includegraphics[width=\columnwidth,trim={0cm 0cm 0cm 10.8cm},clip]{figures/slack-tune-fwd-rev-slice-inv-comparison-2.pdf}}

\end{adjustbox}

\caption{Overview of the TimeStitch proposal.}
\label{Fig:alap-vs-scheduled-technique-compare}

\end{figure}

Current QCs are prototype devices; they are less than 1000 qubits in size and they do not implement fault-tolerant, error correcting codes. These devices suffer from high error rates as noise is introduced during state initialization, gate application, and measurement procedures. 
In addition to errors during operations, qubits are also vulnerable to noise during periods of inactivity.  \emph{Decoherence} error in idling qubits causes state to degrade exponentially 
over time from phase accumulation and amplitude damping. Several near-term applications require critical paths proportional to their circuit size~\cite{cuccaro2004new}, resulting in large qubit idle windows that could lead to decoherence errors.

Our work's fundamental goal is to mitigate both phase accumulation and amplitude damping without needing additional gates in the original circuit.
We present a novel technique to optimize circuits by taking advantage of flexible scheduling within \emph{slack windows}, or periods of qubit idling before its next operation. The benefits of our approach are achieved without extending circuit runtime through either increasing total gate count or introducing circuit partitioning.

Qubit slack may appear trivial in unmapped circuits, but the impact and duration of idling qubits becomes obvious post compilation. Many near-term devices, such as superconducting circuits, feature nearest-neighbor topologies with sparse connectivity across qubits. Unfortunately, many QC applications have communication requirements that do not align well with hardware capabilities, resulting in the insertion of $SWAP$ networks for intra-chip qubit communication. 
These $SWAP$ networks increase the duration of quantum circuits, forcing a significant portion of physical qubit runtime, or time from state initialization until final measurement, to be suspended within slack.

Typical circuit scheduling methods such as ``As Late As Possible'' (ALAP) scheduling, a default approach in IBM Qiskit~\cite{Qiskit}, assume that single-qubit operations are best placed at the end of slack windows. An example of ALAP scheduling is pictured in Fig.~\ref{Fig:alap-vs-scheduled-technique-compare}(a) as an X gate in a dashed box.  
Our proposal reschedules single-qubit gates potentially away from its ALAP default position  to an optimum placement \emph{within the slack window}, also shown in Fig.~\ref{Fig:alap-vs-scheduled-technique-compare}(a). 
The chosen gate placement is deemed optimum by measuring the fidelity at various gate positions in its slack window with a carefully designed tuning circuit.
By choosing the optimal gate position, TimeStitch minimizes the impact of decoherence on qubits caused by dephasing and amplitude damping.

TimeStitch tuning procedures are implemented by intelligently leveraging the reversible nature of quantum computation. First, a tuning circuit starts with a slice of the original circuit up to a target slack window. Next, a window equal to the slack found in the original circuit is placed in the tuning circuit. Finally, the circuit slice is then inverted to ``undo'' previous computation, returning all qubits to their original input state. The approach is thus referred to as a ``slice + inverse" (SI) technique.
Critical to the near-term, TimeStitch tuning employs reversibility without exceeding the machine fidelity limits on circuit depth.
Prior work exploiting reversibility for predicting circuit outcomes builds a concatenated ``forward+reverse'' of a quantum circuit in its entirety, resulting in double the depth of the original circuit~\cite{Patel2021QraftRY}.
Under the reasonable assumption that target circuits are already at or just under machine capacity in terms of critical path length, 
it is possible that the depth of such forward+reverse circuits can far exceed QC frontiers.
This is shown in Fig.~\ref{Fig:alap-vs-scheduled-technique-compare}(b) in blue, and Section~\ref{Reversibility-applied} discusses prior work in greater detail.
Alternatively, our proposal is constrained to specifically target slack windows whose corresponding SI, ``slice + inverse,'' circuits are within the machine limits of circuit depth.
This is shown in green in Fig.~\ref{Fig:alap-vs-scheduled-technique-compare}(b).
Further, we show that even with this constraint we are able to reap most of the potential benefit from our decoherence mitigation approach because it still allows us to target most larger slack windows which both have higher potential for gate position tuning based benefits and create circuit slices of lower gate depth due to the very definition of slack.

\begin{figure}[t]
\begin{adjustbox}{varwidth=0.75\columnwidth,fbox,center}
\includegraphics[width=0.75\columnwidth,trim={0cm 0cm 0cm 0cm},clip]{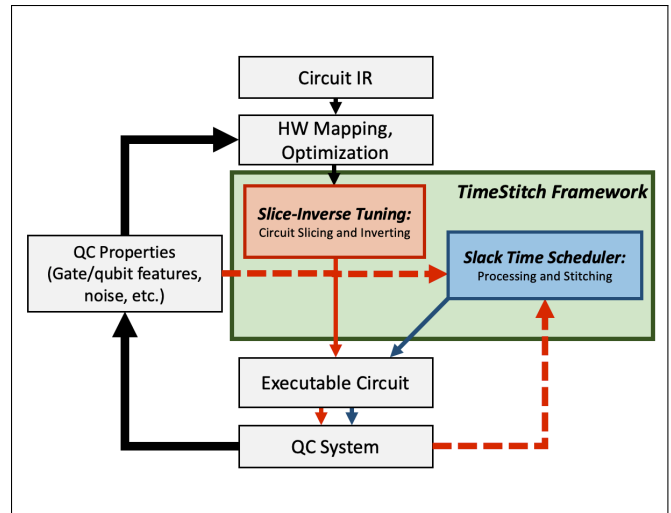}
\centering
\end{adjustbox}
\caption{QC compilation integrated with TimeStitch.}  
\label{Fig:compiler-flow}
\end{figure}

Fig.~\ref{Fig:compiler-flow} depicts a quantum compilation flow that includes the TimeStitch (TS) slack optimizing scheduler which consists of two components. The first component is a module that identifies slack windows within a compiled quantum program and develops slack tuning circuits through circuit slicing and inversion procedures. These circuits are then used to independently optimize single-qubit gate positions within individual slack windows. Compilations and device properties are subject to variation, thus the optimum placements determined by slack tuning will be unique for each circuit and machine at a given period of time. With the optimum scheduling data, TimeStitch creates a final ``stitched'' executable that is an optimized form of the original compiled circuit. 

We note that our proposal to exploit slack in quantum circuits is not limited to our primary error mitigation approach of tuning single-qubit gate positions within slack windows. As a secondary contribution, we show that other techniques for error mitigation, such as dynamical decoupling~\cite{viola1999dynamical}, can also be specifically targeted within these slack windows through our generalized compilation framework. We show that when gate scheduling is intelligently coupled with periodic dynamical decoupling within the TimeStitch framework, the error mitigation techniques compliment each other, resulting in even greater fidelity improvements across a variety of quantum circuits. Finally,
the hallmark theme of exploiting slack in quantum circuits has significant parallels to slack-based optimization in classical computing, such as those at the circuit-level as well as the microarchitecture-level; we dive deeper into these parallels in Section \ref{7-discussion}. 

To summarize, this work makes the following contributions:

\begin{itemize}

\item We observe the creation of slack windows as a result of compilation. To the best of our knowledge, we are the first to identify their potential for optimal quantum gate scheduling.

\item We develop a framework that optimally schedules single qubit gates for mitigating decoherence error that degrades idling qubit state. 

\item To the best of our knowledge, we are the first to exploit quantum reversibility towards gate scheduling, and importantly, in a manner cognizant to device depth limitations. Reversibility enables the mitigation technique to adapt to the unique characteristics across both applications and QCs to provide a solution that is not ``one-size-fits-all.''

\item We design a slack analysis and circuit construction method that analyzes compiled QCs, identifies slack windows, and ``slices'' the original circuit to isolate dependency graphs up until instances of slack. These ``slices'' are then combined with a delay line equal to the corresponding slack window followed by the slice inverse circuit to create a total circuit that evaluates to a ground truth: the slice input state. 

\item We design and implement the TimeStitch Slice+Inverse (TS-SI) slack time scheduler that optimizes the scheduling of single qubit gates within slack. Local optimals are learned during tuning procedures when individual slack windows are searched within slice+inverse circuits (above) to maximize the fidelity of the trivially known ground truth. TS-SI then ``stitches'' a final quantum circuit with optimum placements identified from tuning. During tuning procedures and final circuit creation, the bounds of the original circuit depth are respected as criteria for tuning (TS-SI+C).


\item We implement TimeStitch to suit deployment on real quantum machines and offer insights that can improve the realistic design of future quantum optimization proposals. The framework is evaluated on a variety of benchmark circuits transformed by baseline compilation and TimeStitch Slice+Inverse rescheduling. 
We compare TimeStitch against other scheduling heuristics such as ALAP, ASAP, and Middle, all discussed in Section~\ref{5-m-evalcompare}, and highlight TimeStitch's greater benefits over ``one-size-fits-all'' gate scheduling solutions.

\item We show that our general TimeStitch compilation framework for targeting slack windows can encompass additional error mitigation techniques like periodic dynamic decoupling (DD). Analysis is provided to show that the two approaches can harmonize to create highly-optimized circuits.

\end{itemize}


TimeStitch holds great potential for impact in the area of quantum compiler design as it is the first proposal to exploit optimum scheduling of quantum operators within slack windows. While many existing techniques for mitigating error rely on adding extra gates to the circuit~\cite{viola1999dynamical,giurgica2020digital,das2021adapt} calibrating new gates~\cite{temme2017error}, or extending a circuit's runtime~\cite{murali2020software}, TimeStitch leverages the gates already present in a quantum program in its base form. TimeStitch, however, can be invoked with DD optimization to reap the combined benefits of multiple state-of-art decoherence mitigation techniques. Additionally, a novel aspect of our framework is that unlike previous proposals that employ reversibility through ``forward'' and ``reverse'' circuits~\cite{Patel2021QraftRY}, program duration is not extended either during tuning procedures or in the final rescheduled circuit. This is critical in the near-term where QCs are aggressively pushed to the brink in terms of utilization.

This article proceeds as follows: Section~\ref{2-background} presents background information describing fundamental elements of this study. Section~\ref{3-motivation} details theory related to quantum computing and quantum error mitigation that motivates TimeStitch. Section~\ref{4-solution} describes the design of the TimeStitch framework. Section~\ref{5-methodology} includes the methodology surrounding TimeStitch development and evaluation. Section~\ref{6-evaluation} evaluates TimeStitch with experiments performed on real QCs. Section~\ref{7-discussion} is a discussion of future directions for TimeStitch as well as related work in both the areas of quantum circuit optimization and classical slack exploitation. Section~\ref{8-conclusion} offers conclusions.

\begin{figure}[t]
\centering
\fbox{
\includegraphics[width=0.6\columnwidth,trim={0cm 0cm 0cm 0cm}]{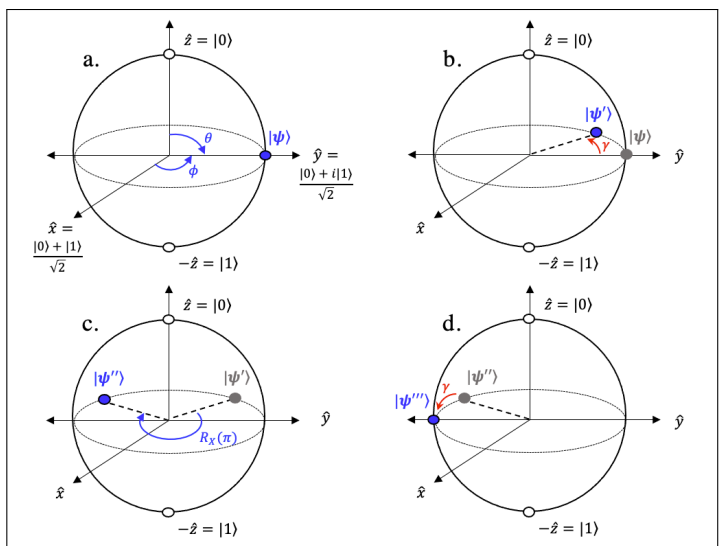}
}
\caption{Phase accumulation mitigation through Hahn spin-echo techniques. (a) A qubit $\ket{\psi}$, prepared with $R_x(\theta=\pi/2)$ and $R_z(\phi=\pi/2)$, rests on the y-axis of the Bloch sphere. (b) As time elapses, the phase of $\ket{\psi}$ decays, and noise in the form of $R_z(\gamma)$ creates the quantum state $\ket{\psi'}$ after a counterclockwise rotation around the z-axis. (c) A $R_x(\pi)$ is applied to the qubit to produce $\ket{\psi''}$, and (d) the effects of dephasing begin to constructively interfere with $\ket{\psi''}$ to produce the phase-coherent state $\ket{\psi'''}$. Another $R_x(\pi)$ pulse restores $\ket{\psi}$ from $\ket{\psi'''}$.}
\label{fig:hahn-echo-example}
\end{figure}

\section{Background}
\label{2-background}
\subsection{Quantum Information and Near-Term QCs}
\label{information-qc}

The basic unit of quantum information is the quantum bit, or qubit. Qubits, unlike classical bits that hold a static values of either 0 or 1, demonstrate superposition in the form of $\ket{\psi} = \alpha\ket{0} + \beta\ket{1}$ where probability amplitudes $\alpha,\beta \in \mathbb{C}$ hold values such that $|\alpha|^2+|\beta|^2=1$. Upon measurement, $\ket{\psi}$ collapses into \textit{either} $\ket{0}$ or $\ket{1}$, effectively becoming a classical bit. A system of $n$ qubits requires $2^n$ amplitudes to describe the state. 

Before measurement, qubits are manipulated with operations, or gates, to modify the quantum state's probability amplitudes. Quantum operations are unitary, and as a result, they are characterized as reversible with the same number of inputs as outputs. Unlike classical computation, there are many non-trivial single-qubit gates such as $R_x(\theta)$ and $R_z(\phi)$ which rotate the state around the x- and z- axis, respectively. An example of $R_x(\theta=\pi/2)$ and $R_z(\phi=\pi/2)$ rotation of qubit visualized with the Bloch Sphere is pictured in Fig.~\ref{fig:hahn-echo-example}(a). Pairs of qubits can be manipulated via multi-qubit interactions. One of the most common of these gates is the two-qubit, controlled-($R_x(\pi)=X$), or $CX$ gate. Together with single qubit gates, $CX$ enables universal quantum computation. There are many choices of basis gate sets specified by the underlying hardware. For more information on the fundamentals of quantum computation we refer to \cite{mike_ike_2020}.

Current QCs, sometimes called Noisy Intermediate Scale Quantum (NISQ) devices, are error prone and less than 100 qubits in size~\cite{preskill2018quantum}. These devices are extremely fragile, and as a result, some of the biggest challenges that limit scalability include limited coherence, gate errors, readout errors, and connectivity. Systematic error is restrictive, but once the error is identified, it can be effectively mitigated or corrected in software. The two primary causes of loss of performance are decoherence and crosstalk. 

Many errors in quantum systems arise from environmental coupling. For example, amplitude damping describes the sporadic loss of energy resulting in the $\ket{1}$ state falling to the $\ket{0}$ state; the rate of this process is described by the device's $T_1$ time. Similarly dephasing, also referred to as phase accumulation or phase damping, details the sporadic change in relative phase and is expressed by the $T_2$ time of the qubit. Both cause qubit state decoherence. Finally, crosstalk refers to error caused by simultaneous execution of gates on nearby qubits. The severity of each type of noise varies per qubit and calibration cycle.

We propose TimeStitch which mitigates different forms of decoherence errors. This is achieved by tuning single-qubit gate positions within idle periods in circuits, Section \ref{3-motivation}.

\subsection{Qubit Idling in Compiled Circuits}
\label{increased-idle-time-compiled-circuits}

\begin{figure}[t]
\centering
\includegraphics[width=0.6\columnwidth,trim={0cm 3.5cm 0cm 3.5cm}]{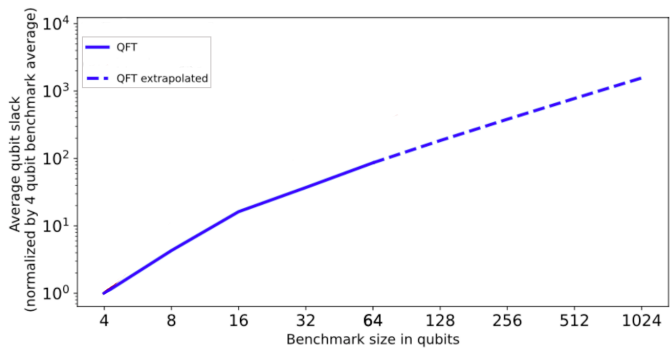}
\caption{Average qubit slack normalized by smallest, four-qubit implementation vs. benchmark size in qubits for QFT benchmark. QFT circuits with 4-64 qubits are mapped to 65 qubit IBM QC, and slack trends are extrapolated to 1024 qubits in anticipation of near-term machines.}
\label{fig:idle-time-bv-qft}
\end{figure}

Qubit runtime during circuit execution is the period spanning the first gate up until measurement. 
During its runtime, a qubit will spend some cycles in computation and others idle  waiting for signals to propagate along a critical path. Idle time is referred to as slack. 

Limited connectivity in near-term devices requires $SWAP$ networks for qubit communication in mapped circuits. As we move towards larger devices, connectivity is anticipated to stay low as architectures such as heavy-hex topologies are expected to be the most favorable to scale superconducting qubit machines~\cite{heavy-hex-blog}. As demonstrated in Fig.~\ref{fig:idle-time-bv-qft}, slack within circuits increases with the number of qubits because of limited qubit-qubit communication. In this plot, the Quantum Fourier Transform (QFT) is mapped to the 65 qubit IBM Q Manhattan quantum machine for 4, 8, 16, 32, and 64 qubit implementations. Maximum optimization is used by the IBM Qiskit~\cite{Qiskit} transpilers. The slack windows that appear in each compiled circuit after the qubit runtime begins are identified, and the total time idling within circuit slack is averaged between all qubits. This average qubit slack for all implementations is then normalized by the smallest slack average corresponding to the four-qubit QFT instance, Fig.~\ref{fig:idle-time-bv-qft} includes a plot (solid line) of the normalized, average qubit idle time total for the 4-64 qubit QFT circuits mapped to the 65 qubit QC. As it is anticipated that nearest-neighbor QCs will scale to thousands of qubits in the near-term, the line detailing average slack is extrapolated (dashed line) from 64 to 1024 qubits to anticipate future technologies. The QFT extrapolated trends show that the circuits have average slack that increases by factors of approximately 1000x at 1024 qubits, demonstrating that the amount of qubit inactivity during its runtime has a direct relationship with circuit size when mapped to near-term hardware. Many quantum algorithms are anticipated to demonstrate this same trend, and regardless of QC technology, QC applications will experience increased circuit slack as algorithms and critical paths scale without substantial parallelization. 

By default, compilation tools tend to schedule single-qubit operations within slack windows for as late as possible (ALAP) meaning that gates will not execute until another operation, typically either a measurement or a two-qubit operation along a critical path, can occur immediately afterwards~\cite{Qiskit}. Scheduling qubit operations for ALAP assists with mitigating noise associated with $T_1$ and $T_2$ decoherence if qubit runtime has not initialized. ALAP execution, however, is not always ideal once a qubit holds state and is more vulnerable to decoherence. Rather than tolerate slack as an unavoidable artifact of compilation and assume ALAP gate defaults, we are motivated to explore theoretical and practical techniques for decoherence mitigation during the periods where qubits idle, as illustrated in Fig. \ref{Fig:alap-vs-scheduled-technique-compare}.

\subsection{Considerations with Applied Reversibility}
\label{Reversibility-applied}

Quantum computation is reversible because quantum operations are unitary. A requirement for a unitary operation, $U$, is that $UU^{-1}=U^{-1}U=(ID)$ where $U^{-1}$ is the operation inverse and $(ID)$ is the identity operation. The identity operation does not evolve qubit state and produces an output equal to the input; it acts as a fixed-duration, ``do nothing'' instruction. As a note, quantum circuit measurement is not reversible 
as it collapses superimposed qubits into a classical bitstring.

A quantum circuit followed by its logical inverse, or a ``forward+reverse'' circuit, thus ideally produces the original or initial state. In QRAFT~\cite{Patel2021QraftRY}, quantum reversibility reduced error in circuits by increasing the likelihood of determining the correct evaluation output. Since the outputs are known as a ground truth for forward+reverse characterization circuits as they are equal to the initial state, noisy QC results can train a machine learning model to discern error attributes for a machine. The model is used to predict true quantum circuit outcomes when circuits are in their forward+reverse form.

While~\cite{Patel2021QraftRY} provides a boost in quantum circuit accuracy, it assumes the ability to successfully run quantum algorithms where critical paths, or depths, are twice that of the original circuit. This may be a reasonable approach for small quantum circuits that terminate well within the bounds of coherence times, but hardware is ideally maximally utilized in practical workloads. Thus, circuits operating at the boundary of machine thresholds may produce unreliable results if executed in their extended forward+reverse form. To avoid observing a noisy distribution, techniques invoking reversibility should consider the duration of the original quantum circuit as bounding criteria.

In this work, quantum reversibility is leveraged by TimeStitch to enable the optimization of single-qubit placement within slack. Unlike~\cite{Patel2021QraftRY} that applies reversibility towards predictive models, we utilize the true output provided by inverting quantum circuits to produce circuit schedules that outperform baseline ALAP compilations. These improvements are achieved without exceeding the critical path criteria either during slack tuning or in the final, optimized circuit. A full description of the TimeStitch framework is found in Section~\ref{4-solution} with details about circuit depth constraints in Section~\ref{Proposed-C}.

\subsection{Spin-echo Error Mitigation: Dynamical Decoupling}
\label{DD_background}

To preserve quantum state without corrective codes, open-loop error mitigation can be applied to refocus signals. An example of this type of correction is dynamical decoupling (DD)~\cite{viola1999dynamical} that ``decouples'' compute qubits from environmental noise. The most elementary form of DD suppresses single-qubit phase accumulation with Hahn spin-echo techniques where $R_x(\pi) = X$ instructions are insert into circuits during runtime. These instructions reverse the impact that dephasing has on quantum state over time. For example, consider a quantum state $\ket{\psi}=\frac{\ket{0}+i\ket{1}}{\sqrt{2}}$. 
 This qubit on the positive y-axis of the Bloch sphere is pictured in Fig.~\ref{fig:hahn-echo-example}(a). Ideally,  $\ket{\psi}$ would hold state information for infinite time, but phase information is highly susceptible to decoherence. In Fig.~\ref{fig:hahn-echo-example}(b), the decay of state by the unknown rotation $R_z(\gamma)$ causes $\ket{\psi}$ to evolve to $\ket{\psi'}$. Hahn spin-echo techniques apply a $R_x(\pi)$ operation to $\ket{\psi'}$ in Fig.~\ref{fig:hahn-echo-example}(c) to mitigate the phase accumulation caused by decoherence, resulting in state $\ket{\psi''}$. The continued dephasing shown in Fig.~\ref{fig:hahn-echo-example}(d) counteracts the original rotation of $R_z(\gamma)$, refocusing phase information to produce qubit $\ket{\psi'''}$. Restoring the original state $\ket{\psi}$, pictured in Fig.~\ref{fig:hahn-echo-example}(a), with phase information intact, is possible with the application of a final $R_x(\pi)$ pulse to $\ket{\psi'''}$. The procedure of inserting $R_x(\pi)R_x(\pi) = XX$ mid-circuit preserves the semantics of the original circuit as 
 $UU^{-1}=(ID)$ where $(ID)$ is the identity operation.

 Many different forms of DD have been proposed~\cite{uhrig2007keeping,quiroz2011quadratic,souza2012robust}, and DD has shown promise on near-term quantum processors~\cite{pokharel2018demonstration,jurcevic2021demonstration,arute2020observation,chen2021exponential,das2021adapt}. 
While DD has considerable potential, the quantum community is still far from widespread implementation due to limitations stemming from non-ideal properties and overheads of the decoupling pulses~\cite{Liu_2013,Souza_2012}. In fact, past work demonstrated that naively implementing DD in a universal manner on idle qubits can result in decreased circuit fidelity~\cite{das2021adapt}. Thus, there is still significant room for DD improvements, both standalone as well as combined with other error mitigation and correction techniques.  

As DD is a leading technique for decoherence mitigation, determining its optimal use in conjunction with TimeStitch was worthy of exploration and resulted in considerable benefits.
 Section~\ref{DD-considerations} includes more discussion motivating integration of DD into the TimeStitch framework.

\section{Theory for Slack Window Optimization}
\label{3-motivation}

\begin{figure}[!t]
\centering
\fbox{\includegraphics[width=0.6\columnwidth,trim={0cm 0cm 0cm 0cm}]{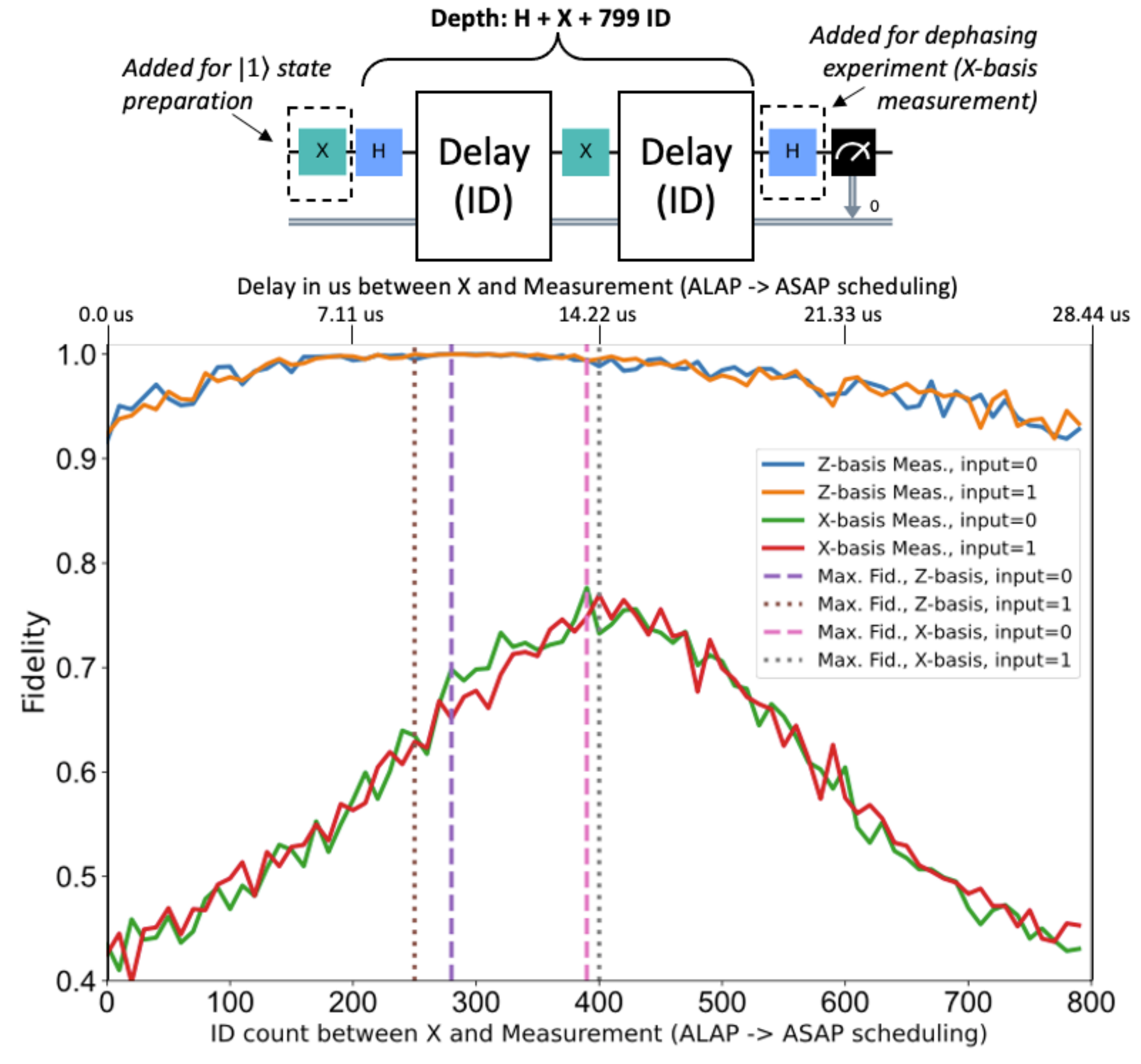}}
\caption{Demonstration of amplitude damping and dephasing correction via Hahn spin-echo techniques. The pictured $H$+$X$+Delay circuit on a single qubit  
tunes $X$ gate placement within a slack window to relate position to state fidelity. Measurement in the Z-basis ($\Ket{0}/\Ket{1}$) captures amplitude information. An $H$ at the circuit end causes an X-basis measurement ($\Ket{+}/\Ket{-}$), capturing phase information. When $X$ is scheduled near the middle of the slack window, the fidelity is maximized. Maximum location for each experiment differs.}
\label{fig:hahn-echo-test}
\end{figure}

\subsection{Tuning Gate Positions for Phase \& Amplitude Errors}
\label{phase-accumulation}

DD techniques employ additional gates to recohere quantum state in the presence of noise. Rather than add gates to a circuit, we are motivated to search for ways to refocus signals using operations already present within the circuit. In the most simple example of how gate placement within slack could influence circuit outcomes, consider the case where a qubit in excited state, $\Ket{\psi_{initial}}=\Ket{1}$, enters an idle period. If the next gate acting on the qubit is an $X$ gate that woud $NOT$ the state of the qubit, lowering it to the ground state, $\Ket{\psi_{final}}=\Ket{0}$, the preferred execution schedule would be as soon as possible (ASAP) to avoid amplitude damping from negatively impacting $\Ket{\psi_{initial}}$ as it idles. Conversely, if $\Ket{\psi_{initial}}=\Ket{0}$ at the beginning of slack, there are advantages to scheduling an upcoming $X$ gate for as late as possible (ALAP) to extend the time the qubit spends in the ground state that is less susceptible to noise. 

Quantum states are often more complex superpositions than those described in the aformented example. For this reason, the circuit in the top of Fig.~\ref{fig:hahn-echo-test} is used as a micro-benchmark to demonstrate the viability for decoherence mitigation via gate rescheduling within slack. An IBM QC was used for this Hahn spin-echo inspired micro-benchmark experiment. The core of the circuit consists of an $H$ gate that puts a qubit into superposition, a slack window artificially created with 799 identity ($ID$), or ``do nothing,'' operations, and an $X$ gate that is tuned within the slack. To tune $X$, the 799 $(ID)$ gates are distributed between two partitions on either side of the $X$ gate that can range from 0 to 799 $(ID)$ gates in size as the $X$ gate sweeps the slack. Additional components included in a select subset of micro-benchmarking experiments are an $X$ gate that prepares the input state $\Ket{1}$ and an $H$ before measurement that allows measurement to be in the X-basis ($\Ket{+}=\frac{\ket{0}+\ket{1}}{\sqrt{2}},\Ket{-}=\frac{\ket{0}-\ket{1}}{\sqrt{2}}$) rather than in the Z-basis ($\Ket{0},\Ket{1}$). These additions are shown in dashed boxes.

The circuit in Fig.~\ref{fig:hahn-echo-test} is inspired by $T_1$ and $T_2$ experiments, but here we do not seek to measure decoherence times. Instead, each of the four versions of the micro-benchmark (input $\Ket{0}$/measurement Z, input $\Ket{1}$/measurement Z, input $\Ket{0}$/measurement X, input $\Ket{1}$/measurement X) has a fixed duration while the $X$ gate position is tuned in search of a maximum fidelity schedule. As a note, each $(ID)$ gate has a duration equal to that of a single $X$ gate on the IBM QCs: approximately 35.56 $ns$. We define fidelity as the Hellinger fidelity between an ideal distribution and the distribution produced from a real QC run. The graph in the lower half of Fig.~\ref{fig:hahn-echo-test} demonstrates that gate placement within slack can influence circuit outcome. Final measurement in the Z-basis ($\Ket{0}/\Ket{1}$) captures information about amplitude damping. An $H$ at the circuit end causes an X-basis measurement ($\Ket{+}/\Ket{-}$), capturing information about qubit phase decoherence. When $X$ is scheduled near the center of the slack window, the fidelity is maximized in all four circuits, although the benefits associated with phase correction were more substantial. This result shows that even though we are not implementing true DD error mitigation, rescheduling \textit{inspired} by Hahn spin-echo techniques can effectively correct both dephasing and amplitude damping error. 

The maximum fidelity schedule for each experiment differs in Fig.~\ref{fig:hahn-echo-test}, suggesting the importance of state and measurement basis for optimum placement. 
In realistic workloads, many variables exist such as variation in single-qubit gate rotation, the qubit that the gate acts on, the slack window state, and the slack duration. The theory alone does not provide a clear prediction of optimum schedule for general use cases, motivating the need for automated solutions that rely on empirical observations, which we pursue by exploiting the quantum property of reversibility.

\subsection{Understanding Real-machine Impact}
\label{real-machine-impact}

\subsubsection{Crosstalk}
\label{crosstalk}

~Crosstalk is the accidental transfer of a qubit's information to surrounding qubits. Two adjacent gates, especially two-qubit interactions, executed simultaneously and within close proximity on nearest-neighbor QCs often experience lower gate fidelity as a result of crosstalk. Because of the severity of crosstalk, software mitigation techniques have been proposed~\cite{murali2020software,ding2020systematic}. 
Studies have shown that single-qubit, single-qubit crosstalk is trivial~\cite{murali2020software}. Thus, the scheduling of single-qubit gates in adjacent slack windows can be tuned independent of one another. Discussion our framework's slack tuning procedures is found in Section~\ref{4-solution}.

\subsubsection{Variation in Qubit Characteristics}
\label{variation-qubit}
~Near-term quantum machines are affected by non-deterministic spatial and temporal variations in their characteristics. For instance, prior work~\cite{Tannu:2019a} observed the prevalence of a wide distribution of machine characteristics with considerable spatial and temporal variation.
From the spatial perspective, they observe the coefficient of variation to be in the range of 30-40\% for $T_1$/$T_2$ coherence times, as well as nearly 75\% for 2-qubit error rates.
From a temporal perspective, they observe more than 2x variation in error rates in terms of day-to-day averages.
  
\begin{figure*}[t]
\fbox{\includegraphics[width=0.95\textwidth,trim={0cm 0cm 0cm 0.35cm},clip]{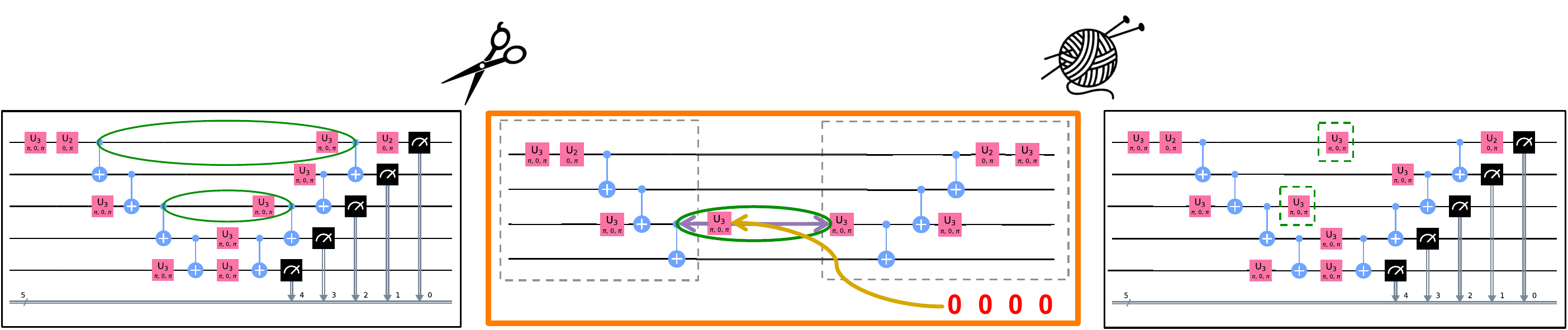}}
\centering
\caption{TimeStitch Framework - (Left:) Circuit is compiled to the target machine and slack windows for tuning are identified throughout the quantum circuit. (Middle:)  In the absence of known circuit outcomes, gate positions are optimized by exploiting quantum reversibility. For each slack window, a circuit slice from circuit start to the slack boundary is constructed and concatenated with a delay line and its inverse. The gate position in the target window is tuned with the goal to make circuit output match the input ($\Ket{00...0}$), implying position of maximum fidelity. (Right:) Tuned gate positions are stitched together to construct an optimized circuit schedule.} 
\label{Fig:Framework_SI_Overview}
\end{figure*}

\subsubsection{State Diversity within Slack Windows}
\label{state-diversity-modivation}

~Each quantum algorithm has a unique objective, resulting in a large amount of state variation during computation, especially within slack. As mentioned, every QC has a distinct noise signature with impact of varying severity depending on an idling qubit's state value.
Unfortunately, certain states are more vulnerable to error. For instance, $\Ket{1}$ is more vulnerable to $T_1$ amplitude dampening than $\Ket{0}$, and $T_2$ dephasing is highly influential to superimposed states such as $\frac{\ket{0}+\ket{1}}{\sqrt{2}}$.

Because of variation within quantum machines, see Section~\ref{variation-qubit}, and how this variation impacts circuits, it is challenging to develop an umbrella benchmark, or a set of benchmarks, for slack tuning that accurately captures unknown state and error attributes seen in real QC execution. Thus, we are motivated to use the circuits and the machines under investigation themselves, building upon the reversible nature of quantum computation, as the basis for slack tuning to accurately capture execution diversity while searching for optimum gate schedules.

\subsection{Considerations for Invoking Dynamical Decoupling}
\label{DD-considerations}

 The $XX$ sequence implements an elementary form of DD that provides Hahn spin-echo correction of phase accumulation. State-of-art DD, however, requires additional gates within the correction sequence because rotation operations around at least two axes are necessary for more robust qubit error decoupling~\cite{souza2012robust}. DD with a single ``universal decoupling'' sequence requires four gates: $R_x(\pi)R_y(\pi)R_x(\pi)R_y(\pi) = XYXY$~\cite{viola1999dynamical}. 
 The universal decoupling sequence adds increased protection to quantum state because $\pi$ rotations about both the x- and y-axis makes the qubit more resilient to environmental noise. Additionally, ~\cite{tyryshkin2010dynamical} analytically shows that $XYXY$ is the superior choice for DD correction of arbitrary quantum states when considering DD sequences containing four gates on two axes.

DD has proven effective at correcting single qubit states and, to a lesser extent, two qubit entangled states in superconducting systems~\cite{pokharel2018demonstration}. 
In addition, DD in the form of the $XX$ sequence to implement Hahn-echo correction has also improved the Quantum Volume (QV) of a real QC in concurrent work~\cite{jurcevic2021demonstration}. Both of these demonstrations, however, cost additional circuit instructions during runtime. 
When inserting DD sequences into a circuit for signal refocusing, the number of additional gates should be carefully considered as gate errors tend to accumulate, potentially destroying the state of the system rather than protecting it from environmental impact~\cite{souza2012robust}. Single-qubit gate errors on superconductors are on average of order $10^{-4}$~\cite{IBMQE,jurcevic2021demonstration}, and although individually small, collective errors can degrade circuit performance, especially as circuits scale on maximally-utilized machines. 

The problem of diminishing quantum circuit outcomes with a naive, universal DD implementation is discussed in related work in framework called ADAPT~\cite{das2021adapt}. ADAPT proposes a clever idea, to evaluate potential DD insertion by transforming target circuits into "decoy" circuits with only Clifford gates.  These novel decoy circuits can then be tractably simulated and selective DD insertion strategies evaluated.
This study shows that in general, there is not a one-size-fits-all solution for DD, but typically adding some DD to a circuit provides improvements. Although impressive performance gains for a small subset of benchmarks are reported using an evaluation metric based on ``total variation distance,'' the benefits of~\cite{das2021adapt} may not be as substantial using more standard metrics such as Hellinger fidelity or probability of success. Additionally, the Clifford approximation used in ADAPT fails to fully model internal states of the circuit, so it is unlikely that the implemented DD is optimum unless the benchmark consists of mostly Clifford operators. In this paper, we propose an alternative that uses circuit slicing and uncomputation to tune DD as well as gate location within paths with high slack using execution on the actual machine.  Our ``slice+inverse'' approach avoids the inaccuracy of the Clifford approximations, but will have some tradeoffs of execution overhead and critical path limitations, as discussed in Sections 5.3 and 4.3. 

In addition to selecting the proper gate sequences and locations for DD within circuits, timing must also be considered so that DD effectively ``unwinds'' error on decaying quantum states. In other words, there must be enough spacing between the execution of gates in the DD sequence to provide corrective benefits~\cite{biercuk2011dynamical,jurcevic2021demonstration}. A commonly implemented form of DD with uniformly-spaced correction gates, such as with $XYXY$, is referred to as periodic DD~\cite{biercuk2011dynamical}. 

For effective application of DD to circuit optimization, it must be deployed through intelligent tuning routines that avoid scenarios of introducing additional gate error that outweighs corrective benefits.
As discussed later in Section \ref{IntTSDD}, intelligent tuning can be especially beneficial when DD is deployed in conjunction with other error mitigation techniques. 
The TimeStitch compilation framework is expanded to incorporate additional decoherence mitigation in the form of periodic DD of $XYXY$, and we empirically tune DD parameters for maximum overall benefit. Empirical details of the methodology are located in Section~\ref{5-m-evalcompare}.

\section{Designing the TimeStitch Framework}
\label{4-solution}

\subsection{Lessons from the Theory}
\label{Theory}
Section~\ref{3-motivation} motivates the need for empirical solutions which are efficient in utilizing the quantum machines.
To do this, these approaches should ideally be backed by robust quantum theory that also take into consideration the abilities of near-term machines. For slack optimization, our proposal for a practical approach is built on the following theoretical lessons:

\circled{1}\ \emph{Prevalence of slack windows}: Section~\ref{increased-idle-time-compiled-circuits} describes that slack windows exist in executable quantum circuits, and their amount and duration are correlated to the size of the quantum circuit targeted for a QC demonstrating limited connectivity between physical qubits.

\circled{2}\ \emph{Adjusting gate positions}: Opportunities for improving the fidelity of quantum circuits exist through adjusting the execution of single-qubit gates from ALAP scheduling to earlier placement within slack windows. A proof-of-concept case study using a micro-benchmark is presented in Section~\ref{phase-accumulation}. 

\circled{3}\ \emph{Optimal positioning}: Optimal gate scheduling within a slack window depends on gate and qubit characteristics along with input qubit state to the slack window. The vast space of these parameters on real machines, Section~\ref{real-machine-impact}, suggests that offline machine characterization on test inputs and circuits is insufficient and impractical for finding optimal gate positions for general use cases.

\circled{4}\ \emph{Reversibility}: Although we cannot predict the outcome of a quantum circuit execution, Section~\ref{Reversibility-applied} describes that quantum reversibility can be used to provide a ground truth. We are motivated to apply reversibility to learn properties of quantum circuits and machines within windows of slack to implement application-specific decoherence mitigation.

\circled{5}\ \emph{Impact of single-qubit crosstalk}: Minimal impact of single-qubit crosstalk, Section~\ref{crosstalk}, means that the single-qubit gate position in each slack window can be optimally tuned independent of those in other slack windows.

\circled{6}\ \emph{Synergistic TS deployment with prior art}: Decoherence mitigation techniques such as DD exist and are shown to be effective, but these solutions must be carefully implemented so that the operational characteristics unique to a circuit and machine pairing are considered. 
In the context of this work, a slack window must be of a minimum duration to provide adequate spacing between DD sequence gates within the window so that DD is effective and DD gate errors are trivial.
Integrating the proposed TS technique with DD involves re-evaluating the best windows to incorporate DD, as slack windows are divided when gate positions are adjusted. 
Details are found in Section~\ref{DD_background} and~\ref{DD-considerations}.

\subsection{A Practical ``Slice + Inverse'' Approach}
\label{Proposed}

The practical TimeStitch approach leverages the quantum phenomenon of reversibility to adjust the execution timing for single-qubit gates within slack windows through the process of circuit slicing and inverting (SI). 
An overview of the framework is shown in Fig.~\ref{Fig:Framework_SI_Overview} and is discussed below.

\subsubsection{Baseline Compilation of the Quantum Circuit}
\label{GT-compile}
The TimeStitch framework begins with a quantum circuit compiled from a device-independent intermediate representation (IR) into machine-ready code. 
The baseline circuit, methodology discussed in Section \ref{Infra}, appears in the left circuit of Fig.~\ref{Fig:Framework_SI_Overview}.

\subsubsection{Identifying Slack Windows}
\label{id-slack}
The TimeStitch framework identifies quantum circuit slack windows after baseline compilation. The identification procedure requires traversing the components of a quantum circuit that implements default ALAP scheduling from end to end. During this procedure, slack windows are found, their durations are calculated using gate timing data collected from the QC. A subset of windows are identified that contain single-qubit operators eligible for rescheduling within slack.
Two such windows are circled in green in the left circuit of Fig.~\ref{Fig:Framework_SI_Overview}. As a note, we do not consider the time before the first operation on a qubit as slack since the qubit is uninitialized and its runtime has not begun.

\subsubsection{Generation of Slice+Inverse Calibration Circuits}
\label{5-generation-SI-Circuits}

 From the identified slack, calibration circuits consisting of sliced partitions of the original circuit and their corresponding inverses are generated. 
Tuning experiments determine optimal schedules for single-qubit gates that are suitable candidates depending on criteria set by the depth of the original circuit targeted for optimization. Setting a criteria for tuning circuit depth is critical for ensuring that calibration procedures to not exceed the frontier of the QC. 

We restate that our goal is to find the optimum gate position within each slack window to boost overall circuit fidelity. 
We employ the property of quantum reversibility, described in Section \ref{Reversibility-applied}, to determine optimum single-qubit execution times within circuit slack to mitigate decoherence and thus maximize the fidelity at the end of the slack. 

For each eligible slack window, for example, the window circled in green in the center circuit of Fig.~\ref{Fig:Framework_SI_Overview}, a circuit slice is constructed, terminating at the end point of the particular window. Implementation-wise, this circuit slice is simply a subcircuit of the original circuit consisting of the dependency graph up until the end of the slack window. Qubit mapping of the original circuit is preserved in the subcircuit, thus the output of the circuit slice emulates the activity of the circuit up to the end-point of the slack window. The inverse of the circuit slice is then constructed and concatenated at the endpoint of the slice. An example slice and its inverse is shown in dashed boxes in the center circuit Fig.~\ref{Fig:Framework_SI_Overview};  we refer to this as a ``slice + inverse'' (SI) circuit.  Measurement operations, not shown in Fig.~\ref{Fig:Framework_SI_Overview}, are added to the end of the SI circuit. In an ideal, noise-free setting, the concatenation of a slice and inverse would produce the slice's input as the output of the inverse because of quantum reversibility. In a realistic, noisy setting, our goal is then transformed to tuning the gate position in the slack window so that the probability of achieving the slice input state as the output of the total concatenated circuit is maximized. This is equivalent to maximizing the circuit fidelity of the original slice under the reasonable assumption that noise impact on the slice and its inverse are well correlated.

The input state to the slice is trivially known if the slice is constructed from the start of the entire quantum circuit; it is the ground state or $\Ket{00...0}$. As a result, the target output of the SI circuit is also the ground state $\Ket{00...0}$ as shown in red in the center of Fig.~\ref{Fig:Framework_SI_Overview}. Since input states, gates, and noise characteristics all influence the optimal gate position, each slack window must be sliced individually. TimeStitch creates the required, unique SI circuits in an automated manner, and the total number of SI tuning circuits for an input circuit is equal to the number of identified slack windows with single-qubit operations.  

As a note, that the maximum depth of an SI circuit is approximately twice the depth of the original circuit if a slice extends to near the circuit's end point. In the approach discussed in this Section, we do not limit the depth of the SI circuits; constraining the depth is discussed in Section \ref{Proposed-C}.

\subsubsection{Optimal Gate Placement in Slack Windows}

Optimal gate placement within slack windows is determined by locating the position within each SI slack window where the ground truth $\Ket{00...0}$ is maximized. A variety of search strategies can be employed to find the local slack optimum, but optimizing the window search is orthogonal to this work.  Each window is optimized independently with its corresponding SI circuit, and resolution of the search can be selected based on the availability of quantum machine resources. In other words, although tuning overhead is manageable and worthwhile for notable quantum circuit fidelity gains on real hardware, if a user is limited by the number of available quantum circuit runs, each calibration circuit can be more coarsely searched. 

\subsubsection{Stitching the Per-window Positions Together}
\label{GT-Stitch}
After the optimal schedules are estimated via SI tuning, the local schedules are stitched together to form a composite, rescheduled circuit, as pictured in the right of Fig.~\ref{Fig:Framework_SI_Overview}.

\begin{figure}[t]
\fbox{\includegraphics[width=0.6\columnwidth,trim={0.8cm 1.5cm 0.2cm 1cm}]{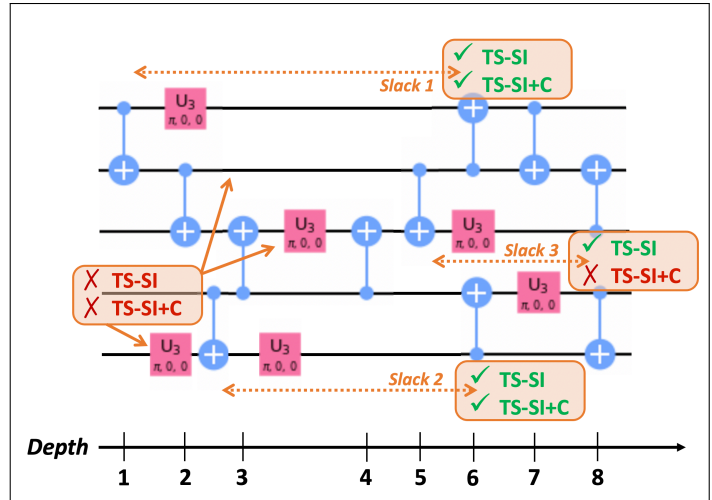}}
\centering 
\caption{Eligible and ineligible circuit locations for TS-SI and TS-SI+C tuning. Slack windows 1, 2, and 3 are eligible for TS-SI as they all have tunable single-qubit operations. However, only windows 1 and 2 satisfy the depth criteria and are thus tuned by TS-SI+C.}  
\label{Fig:slack-eligibility}
\end{figure}

\subsection{Constraining by Circuit Depth}
\label{Proposed-C}

The total number of SI tuning circuits is equal to the number of slack instances that contain tunable single-qubit gates. However, some of these SI circuits, such as those that cover slack appearing at the end of the input circuit, may have a depth exceeding that of the original circuit.
This is because the SI tuning circuit will have a depth twice that of the subcircuit slice leading up to the slack window, as seen in the center of Fig.~\ref{Fig:Framework_SI_Overview}. Past work~\cite{Patel2021QraftRY} leveraging reversibility does not take circuit depth increase into consideration, but not doing so could potentially push beyond the frontier of the targeted device.
The intuitive reasoning is that in the near term we are likely to be executing quantum applications that are already at the brink of a QC's capability in terms of the machine's critical circuit depth.
Building tuning circuits beyond this critical depth can be detrimental to optimizing the original circuit because it may provide false optimum schedules that are distorted by noise. To be mindful of the limitations of gate error and decoherence in current QC hardware, TS-SI can be run using the depth, or critical path, of the original circuit as bounding criteria. This version of TimeStitch, illustrated earlier in Fig.\ref{Fig:alap-vs-scheduled-technique-compare}(b), is known as Slice+Inverse+Criteria (TS-SI+C). 

Here, we focus only on the two-qubit operations along the critical path as a measure of depth.
Two-qubit operations dominate in terms of influence on program output because on average, their error rates and duration are >10x of a single-qubit operation~\cite{jurcevic2021demonstration}. 
This is particularly favorable for TimeStitch because large slack windows have two-qubit depth that are often considerably lower than the critical depth, which is  why large slack window exists.
Moreover, large slack windows are likely to provide substantial benefits due to the wider space for gate position tuning.  

With TS-SI+C, depth is calculated for each of the SI tuning circuits. Those having a depth less than or equal to the depth of the original circuit are marked for use during TS-SI+C slack window gate position tuning. All untuned slack windows maintain default ALAP scheduling. 
Examples of circuit locations eligible and ineligible for TS-SI and TS-SI+C tuning with TimeStitch optimization are pictured in Fig.~\ref{Fig:slack-eligibility}. In the original compiled circuit used as TimeStitch input, slack windows 1, 2, and 3 are eligible for TS-SI as they all have tunable single-qubit operations. However, only slack windows 1 and 2 satisfy the depth criteria since their corresponding SI circuits are of lower depth than the original circuit. Thus, only slack windows 1 and 2 are tuned by TS-SI+C. There are many other locations in the circuit, such as slack windows without single-qubit gates or periods before qubit runtime begins that are ineligible for slack tuning. This is also illustrated in Fig.~\ref{Fig:slack-eligibility}.

\subsection{Integrating TS with Dynamical Decoupling}
\label{IntTSDD}
As mentioned in Section~\ref{DD-considerations}, DD sequence gates are spread within an idle window with adequate spacing between gates to provide maximal decoherence mitigation. Additionally, we wish to provide maximal correction benefits with DD without increasing the number of gates to the point where gate errors accumulate and degrade the state of the system~\cite{souza2012robust}. Slack duration in compiled quantum circuits are prone to a vast amount of variation across applications and QCs, and implementing TimeStitch optimization results in the challenge of cutting large idle windows into smaller segments of size that is unknown before SI tuning procedures. DD implementation is highly dependent on window size, so this presents a challenge for employing DD to generate a maximally TimeStitch-scheduled circuit. Thus, we are motivated by the potential of DD to empirically develop a heuristic that generalizes DD to a vast set of use cases. This heuristic will be based on the periodic $XYXY$ sequence and serves as an additional optimization benefit of the TimeStitch framework. More information is found in Section~\ref{5-m-evalcompare}.

\section{Methodology}
\label{5-methodology}

\subsection{Evaluation Effort - Quantities \& Constraints}

We perform all our experiments on actual IBM superconducting quantum machines to faithfully capture true device characteristics.
 Our evaluations encompass roughly 2,500 quantum jobs to the cloud, comprising of over 600,000 circuits, with confidence built on a total of over 4,000,000,000 QC executions.
 Out of these, we show results for 10 applications, each paired to a target machine satisfying the following criteria: a) consistent machine availability, b) non-negligible probability of the correct application output during baseline evaluation, c) limited variability in correct output probabilities, and d) maximizing machine qubit utilization while respecting the previous constraints. 

\subsection{Circuits for Evaluation}
\label{5-m-circ}
Our benchmarks are representative of real-world usecases, described here and in Table \ref{tab:method}. 
Due to limitaions on circuit width because of machine size and depth because of coherence times on available near-term QCs, benchmarks that included 6 qubits or fewer and of shorter duration were included in TimeStitch evaluation. 
Brief descriptions of the benchmarks used in our study are as follows.

\begin{itemize}

\item\emph{Quantum Fourier Transform:} 
 QFT is a circuit used as a building block for applications such as Shor's algorithm for quantum factoring~\cite{shor1999polynomial} and phase estimation.
It converts a quantum state from the computational basis to the Fourier basis through the introduction of phase. 
QFT was constructed for 4 and 5 qubits~\cite{mike_ike_2020}.

\item \emph{Quantum Approximate Optimization Algorithm:} 
QAOA \cite{farhi2014quantum} is a variational quantum-classical algorithm to solve combinatorial optimization problems.
QAOA is implemented atop a parameterized circuit called an ansatz. We use one instance of a hardware efficient QAOA ansatz, and its solution is simple to predict when solving MAXCUT on a ``ring of disagrees'' graph structure. 
We use QAOA ansatz constructed for 4 and 6 qubits.

\item \emph{Variational Quantum Eigensolver:}
VQE~\cite{peruzzo2014variational} is a hybrid algorithm like QAOA and is used to variationally find the lowest eigenvalue of a given problem matrix
by computing a difficult cost function on the QPU and feeding this value into an optimization routine running on a CPU. 
We implement VQE on a hardware-efficient SU2 ansatz~\cite{IBM-SU2} and use one instance as the benchmark.
We construct the ansatz for 4 qubits and 6 qubits.

\item \emph{Gibbs State Prep:}
The preparation of Gibbs state has applications in quantum simulation, optimization, and machine learning.
We take a VarQITE ansatz based approach to create the Gibbs state~\cite{Zoufal_2021}.
We use 5 qubits for the Gibbs circuit. 

\item \emph{Quantum Repetition Code Encoder:}
Error correcting codes are the means by which fault-tolerant quantum computers are able to execute arbitrarily long programs. 
Many such codes have been developed that make multiple tradeoffs~\cite{divincenzo1996fault, calderbank1998quantum, fowler2012surface}.
Here, we target a error correcting repetition code encoder whose effect is to distribute the quantum information in the initial state across an entangled N-party logical state. 
This introduces redundancy to the encoding that can be exploited for error detection~\cite{Roffe_2019}.
We use an encoder targeting 5 qubits.

\item \emph{Greenberger–Horne–Zeilinger (GHZ) State Prep:} 
GHZ state~\cite{greenberger1989going} generation is 
a non-traditional benchmark, but  
useful as many complex quantum algorithms begin by entangling all qubits before computation in a state preparation process. 
In this benchmark, all qubits are first fully entangled before $X$ gates swap the $\Ket{0}$ and $\Ket{1}$ probability amplitudes. 
Finally, qubits are unentangled to restore the input state. 
GHZ was implemented for 5 qubits.

\item \emph{Ripple Carry Adder:}
Adders are a critical logic building block for quantum logic such as in Shor's algorithm for quantum factoring~\cite{shor1999polynomial}. 
We implemented a linear-depth, two-bit ripple-carry adder quantum circuit that uses 6 qubits~\cite{cuccaro2004new}.

\end{itemize}

\subsection{Infrastructure:}
\label{Infra}
TimeStitch is implemented as a compilation pass that performs schedule optimization on top of a highest-baseline compilation of Qiskit Terra 0.16.4 to map and optimize for the IBM machines~\cite{Qiskit}. 
We distribute across 5 quantum devices: Casablanca (7 qubits), Jakarta (7 qubits), Guadalupe (16 qubits), Toronto (27 qubits), and Sydney (27 qubits).
Machine details on the IBM Quantum Systems page~\cite{IBMQS}.

The IBM QCs are accessed via the quantum cloud. Resources are shared among hundrerds of thousands of users running more than two billion experiments per day~\cite{use-blog}, causing the queue time to service a quantum experiment request to vary significantly. As a result, an efficient means to utilize the cloud QCs is to maximize batches, or jobs, sent to a IBM QC. Jobs are treated as a single task, allowing for for the sequential processing and combined results of multiple circuits. A single quantum job of our target QCs can execute a batch of up to 900 circuits.

To keep the tuning overhead manageable, restricting calibration within a single job is key as the job runtime is more predictable and often significantly shorter than queue time in the IBM quantum cloud~\cite{gokul-res-manage-2021}. We use utilize this entire batch across the tuning of gate positions for different slack windows.
Thus each slack window gets $N = \frac{900}{\# SW}$ circuit slots for tuning, and the resolution of each window's gate position sweep is $R = \frac{N}{SW_{length}}$.

\begin{table}[]
\resizebox{.6\columnwidth}{!}{%
\begin{tabular}{|l|l|l|l|l|l|l|}
\hline
\textbf{Bench} &
  \textbf{Q} &
  \textbf{D} &
  \textbf{Output} &
  \textbf{\begin{tabular}[c]{@{}l@{}}\# SW\\ /Cons.\end{tabular}} &
  \textbf{\begin{tabular}[c]{@{}l@{}}Avg.SW\\ (1e-6 s)\end{tabular}} &
  \textbf{Dev} \\ \hline
\textbf{QAOA}  & 4 & 15 & \begin{tabular}[c]{@{}l@{}}0101 +\\ 1010\end{tabular}          & 10/5  & 0.85 & Guad \\ \hline
\textbf{QAOA}  & 6 & 84 & \begin{tabular}[c]{@{}l@{}}101000 +\\ 111101\end{tabular}      & 19/10 & 0.91 & Jak  \\ \hline
\textbf{Gibbs} & 5 & 12 & \begin{tabular}[c]{@{}l@{}}0000+0101+\\ 1010+1111\end{tabular} & 3/2   & 1.17 & Jak  \\ \hline
\textbf{QFT}   & 4 & 29 & 1010                                                           & 15/8  & 1.25 & Tor  \\ \hline
\textbf{GHZ}   & 5 & 8  & 10101                                                          & 3/3   & 1.48 & Syd  \\ \hline
\textbf{VQE}   & 4 & 63 & 0111                                                           & 18/10 & 1.67 & Guad \\ \hline
\textbf{QFT}   & 5 & 39 & 00101                                                          & 25/12 & 1.83 & Tor  \\ \hline
\textbf{QEC}   & 5 & 26 & \begin{tabular}[c]{@{}l@{}}00000 +\\ 01011\end{tabular}        & 4/4   & 2.41 & Cas  \\ \hline
\textbf{Adder} & 6 & 64 & \begin{tabular}[c]{@{}l@{}}000110\\ (1+0)\end{tabular}         & 64/8  & 3.02 & Syd  \\ \hline
\textbf{VQE}   & 6 & 51 & 111111                                                         & 13/5  & 3.99 & Cas  \\ \hline
\end{tabular}%
}
\caption{Benchmarks and their characteristics, sorted by Avg.SW. Q: Number of application qubits, D: Circuit depth in CXs, Output: Application outputs, \# SW/Cons.: Number of slack windows / slack windows targeted under depth constraint, Avg.SW: Average window size, Dev: Target machine.}
\label{tab:method}
\end{table}

\label{5-m-metrics}

The benefits of our proposal on these circuits is evaluated on the \emph{Probability of Success (POS)} metric which is the ratio of a number of error-free trials to the total number of trials - a common metric for evaluating quantum optimization.

\subsection{Evaluation Comparisons:}
\label{5-m-evalcompare}
To analyze the effectiveness of TimeStitch, its performance was compared to other universal gate scheduling techniques as well as alternative error mitigation techniques targeted towards slack. The set used in comparision to TimeStitch are described below. 

\begin{itemize}
\item \emph{As Late As Possible (ALAP):} ALAP is the default scheduling technique implemented in the Qiskit compiler.
All gates appearing in slack windows will be executed at the end immediately before the next two-qubit gate that acts as the slack end boundary. 
As ALAP is the default compilation setting of the Qiskit compiler, it acts as the baseline comparing the TimeStitch framework to other scheduling and optimization techniques described in this Section.

\item \emph{As Soon As Possible (ASAP):} ASAP forces all gates appearing in slack windows to be executed immediately after the two-qubit gate that acts as the slack beginning boundary.

\item \emph{Middle:} Middle scheduling is a naive scheduling technique that executes all single qubits within slack at the center of their slack windows.

\item \emph{TimeStitch with Circuit Slice+Inverse Tuning (TS-SI):} TS-SI corresponds to the design from Section~\ref{Proposed}. 
The TS-SI form of the proposed optimization framework is ``unconstrained'' and invokes slack tuning on all circuit windows using slice+inverse calibration circuits, regardless of circuit depth. The final optimized output circuit, however, is of depth equal to that of the original circuit.

\item \emph{TimeStitch with Circuit Slice+Inverse Tuning, plus Criteria (TS-SI+C):} TS-SI+C corresponds to the design from Section~\ref{Proposed-C}. 
TS-SI+C is a practical approach as it respects the critical path of the original circuit. Windows are not tuned if their slice+inverse circuit exceeds the depth of the original circuit. Windows that are not tuned because of depth criteria violation maintain default ALAP scheduling.

\item \emph{Dynamic Decoupling (DD):} In this article's implementation of DD, universal $XYXY$ decoupling appears in slack windows. A single round is added to the window if it fits within the window duration. The gates are evenly spread across the window to create a periodic sequence. 

\item \emph{Dynamic Decoupling w/ Heuristic (DD(H)):} DD(H) is similar to above, but DD is only added if a heuristic threshold inspired by~\cite{jurcevic2021demonstration,souza2012robust,tyryshkin2010dynamical} is met to allow for adequate spacing between DD pulses. The inspiration for our DD heuristic is discussed further in Section \ref{DD-considerations}. For our evaluation, the empirically found duration threshold is that the slack duration should be greater than or equal to four times the DD sequence duration.

\item \emph{Integrated TimeStitch and Dynamic Decoupling (TS+DD):} TS+DD is the combined deployment of TS and DD wherein DD is inserted into slack windows created post gate scheduling, as discussed in Section \ref{IntTSDD}.

\item \emph{Integrated TimeStitch and Dynamic Decoupling w/ Heuristic (TS+DD(H)):} TS+DD(H) is similar to above, but also incorporates DD insertion according to the heuristic slack threshold.

\end{itemize}

\begin{figure}[t]
{\includegraphics[width=\columnwidth,trim={0cm 0cm 0cm 0cm},clip]{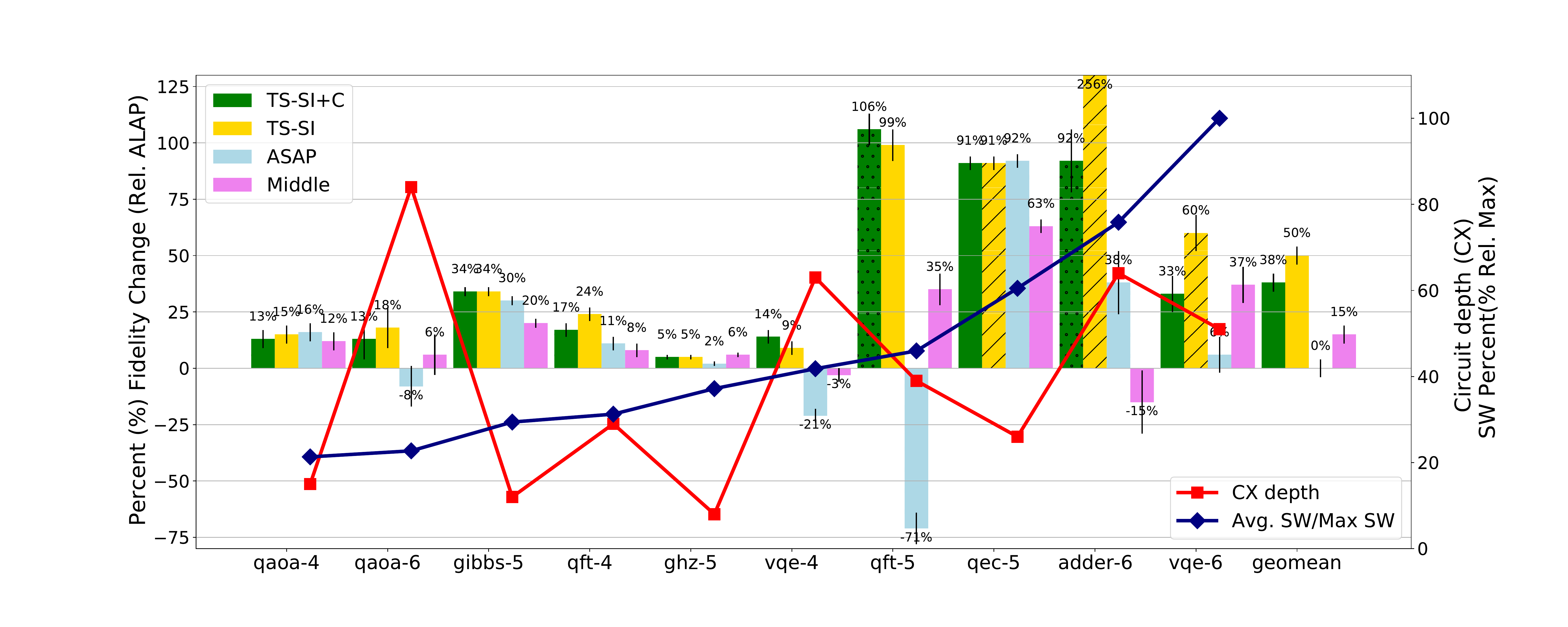}}
\centering
\caption{POS benefits of different approaches over the ALAP baseline. TS-SI benefits are highest at 50\% mean improvement over the baseline. TS-SI+C provides a 38\% improvement. ASAP and Middle provide negligible and lower benefits on average, respectively, with detrimental individual outcomes on some individual benchmarks. In general, TS benefits increase as slack windows grow. These results were generated with real QC experiments}  
\label{Fig:Result-POS}
\end{figure}

\section{Evaluation}
\label{6-evaluation}

\subsection{Probability of Success}
\label{eval-pos}

In Fig.~\ref{Fig:Result-POS} we show benefits in terms of POS improvements relative to the ALAP baseline.
Benefits shown are in terms of the relative increase in POS.
for TimeStitch (TS), we show results for TS-SI+C and TS-SI.
We also show comparisons to ASAP and Middle. 
All are detailed in Section \ref{5-m-evalcompare}.
Applications are ordered by their relative average slack window sizes, and in general, larger slack windows provided greater benefits.

TS-SI achieves a 50\% POS geometric mean improvement, clearly showing the efficiency of the slice+inverse technique in meeting the ideal improvement target.
TS-SI+C  constrains the slice+inverse technique, so that no SI tuning circuit exceeds the gate depth of the original circuit.
Even with this constraint, a mean 38\% improvement is obtained, indicating that even under constraint, multiple critical slack windows can be tuned for significant benefits.
In comparison, ASAP and Middle achieve no benefit and 15\% mean POS improvement respectively, and observe POS reductions or negligible benefits across many individual benchmarks. While showing some promise, both ASAP and Middle occasionally produce slowdowns on some benchmarks or minimal benefits on others. We can conclude that they are not optimal scheduling solutions.
A ``one-size-fits-all" approach does not maximize benefits, clear quantitative motivation that specifically tuned gate positions are preferred.
These highlight the benefits of tuning single-qubit gates within slack windows, especially with the practical TS approach that harnesses quantum reversibility in a novel manner for observable quantum circuit gains. 

Over the two TS techniques, per-application improvements vary from 5\% to 256\%. We reason that this variation is caused by the number and sizes of the slack windows, the criticality of the slack window to the circuit, impact of specific gate errors on application fidelity, the input state vectors, as well as general noise characteristics of the machine.
Table.~\ref{tab:method} provides some compiled circuit details.
As a note, TS optimizes circuit schedule within slack without increasing the depth or duration of the benchmark.

In Fig.~\ref{Fig:Result-POS}, plots of $CX$ depth and average slack window size relative to the maximum average slack window are included.
It is clear that benefits increase with greater average slack window size.
This is important because slack durations will increase as applications become more scale and require more $SWAP$s for qubit communication, as discussed in Section~\ref{increased-idle-time-compiled-circuits}. 
We add error bars to the graphs to indicate variation in relative POS benefits from a 1\% change to the application's POS.
For applications with lower baseline fidelity like the Adder (around 10\%), these error bars are longer, but POS benefits are considerable irrespective. This is important as on near-term devices, it is critical in the near to improve the execution of applications with borderline or less than acceptable circuit fidelity.
Some variation is expected across runs depending on the particular run's machine characteristics and calibration.

\begin{figure}[t]
{\includegraphics[width=0.6\columnwidth]{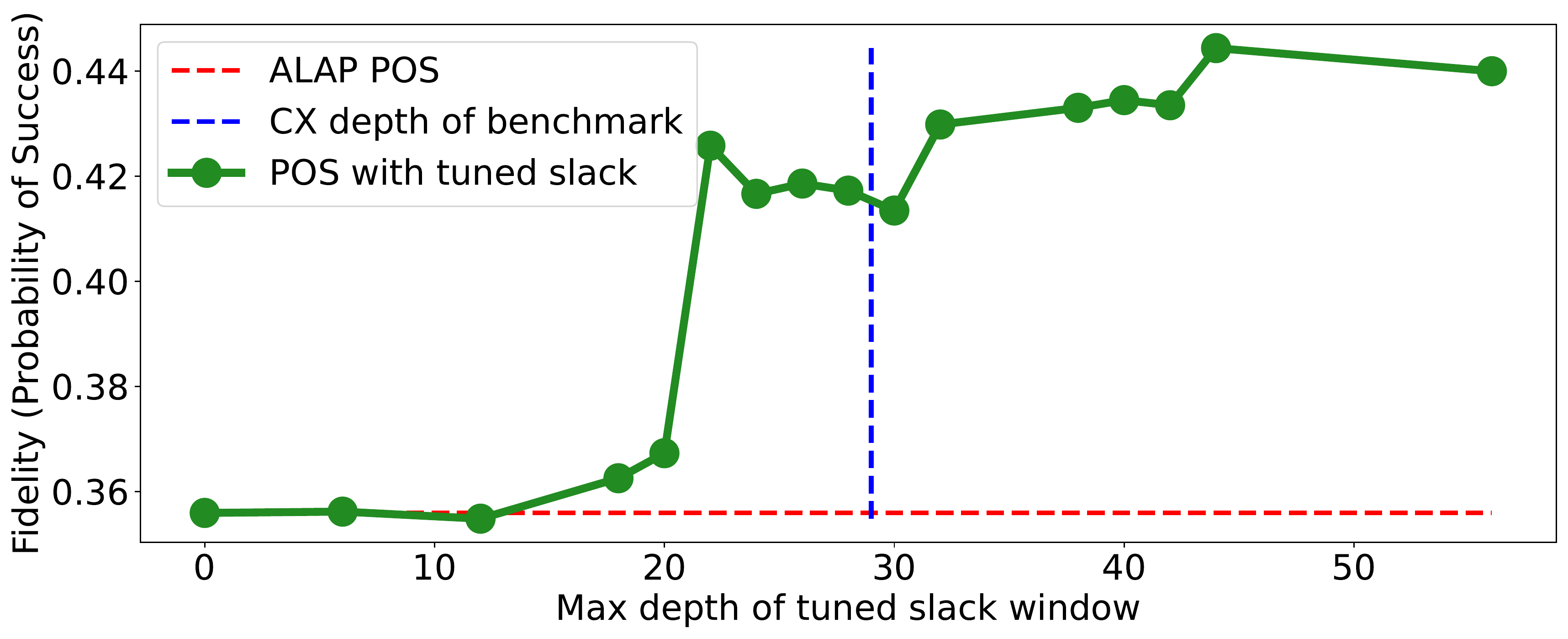}}
\centering
\caption{Threshold sensitivity of window tuning for QFT-4. The red line represents the ALAP POS, the blue line indicates the circuit depth criteria used for TS-SI+C, and the green line describes change in POS as the number of SI tuned slack windows increases.}
\label{Fig:TS-sensitivity}
\end{figure}

\subsection{Depth Threshold Sensitivity}

Here, we fix the criteria of TS-SI+C to respect the depth of the original circuit undergoing optimization. Reasoning for this design choice was the assumption that QCs in the near term will implement high utilization with respect to algorithms that they will run. Therefore, it is essential to prohibit tuning processes with circuit slicing and inverting mechanisms that push beyond the frontier of machine capabilities in terms of coherence time and gate error.  

In Section \ref{Proposed-C} we motivated the need for restricting the SI tuning circuit depth to the depth of the original circuit.
This evaluation involves sweeping through varying limiting thresholds for the depth of an SI circuit from 0 or no slack windows tuned to 2x the original circuit depth, or all slack windows are tunable, equivalent to the unconstrained approach in Section \ref{Proposed}). Fig.~\ref{Fig:TS-sensitivity} shows the POS for the QFT-4 circuit for these different thresholds. 
The baseline ALAP POS (red line) as well as the depth of the original circuit (blue line) are also shown for comparison. 
Tuning windows are ordered by size in Fig.~\ref{Fig:TS-sensitivity}, and TS criteria is satisfied in the first 8 of the 15 windows. 
Adjusting the target depth threshold of slack tuning can influence the POS. 
With original circuit depth as the limiting constraint, TS is able to improve the fidelity from 35\% to 42\%.
If machine robustness allowed an unlimited, or at least a 2x, tuning circuit depth, perhaps in the case that the target circuit fell well below coherence bounds, POS jumps to 45\%.
Depth thresholds can be set based on the machine-application fit.
Note that the experimental results suffer from some variation effects of the real machine hence we do not see a strictly monotonically increasing curve.

\begin{figure}[t]
{\includegraphics[width=.6\columnwidth]{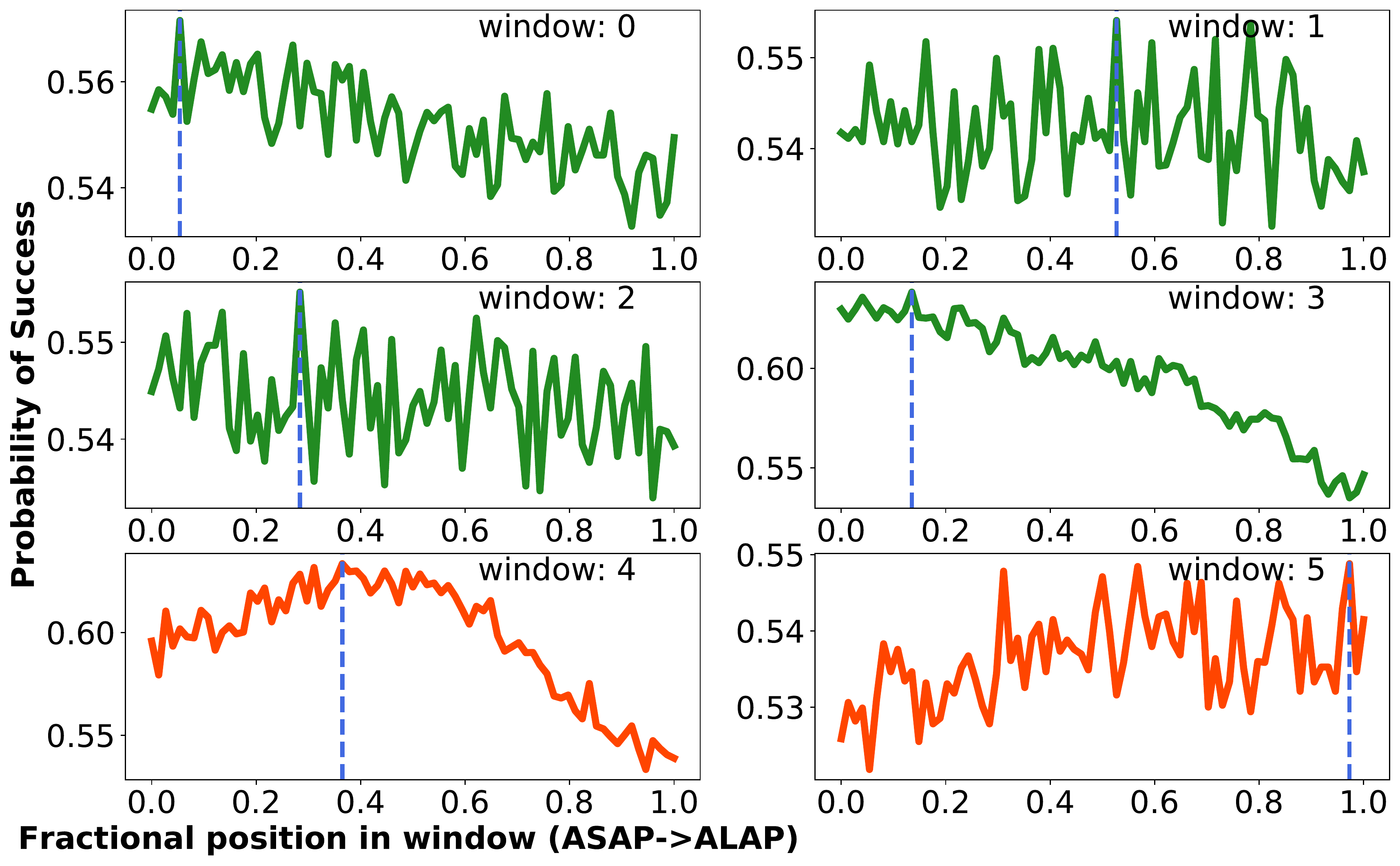}}
\centering
\caption{Slack windows of QEC-5, showing range of POS achieved by tuning each window. Green windows are selected under TS-SI+C while red are rejected. Note that maximum POS for each window differs.}  
\label{Fig:QEC-Slack}
\end{figure}

\subsection{Comparing Slack Windows}

In Fig.~\ref{Fig:QEC-Slack} we show the slack windows of the QEC-5 application, compiled to a 5 qubit machine, as a case study. The change in QEC-5 benchmark POS is plotted as gate positions within an isolated slack window are varied from ASAP (left) to ALAP (right).
The windows that are suited to the depth constraint imposed by TS-SI+C are shown in green while the others are in red.

First, it is clear that there are non-negligible POS variations in four out of the six windows and all windows have different optimal gate positions.
Second, among the green windows, there is considerable benefit in moving to ASAP for window:3. 
Third, among the red windows, there is considerable benefit for window:4 near the middle.
With the TS framework, all local optimums are stitched together to create a final, schedule-optimized circuit.
Thus, the benefits of TS-SI+C are considerable over ALAP baseline, and relaxing constraints with TS-SI can produce even greater benefits.

\subsection{Leveraging Dynamical Decoupling}
Dynamical Decoupling (DD) is an established error mitigation technique with similar inspirations as TS.
We observe that the mitigation effects of the DD and TS approaches interfere constructively and thus the two can be deployed in a synergistic manner.
The benefits can be further improved via intelligent tuning by means of a DD insertion heuristic threshold discussed in Section \ref{IntTSDD}.

\begin{figure}[t]
\centering
\includegraphics[width=\columnwidth,trim={0cm 0cm 0cm 0cm}]{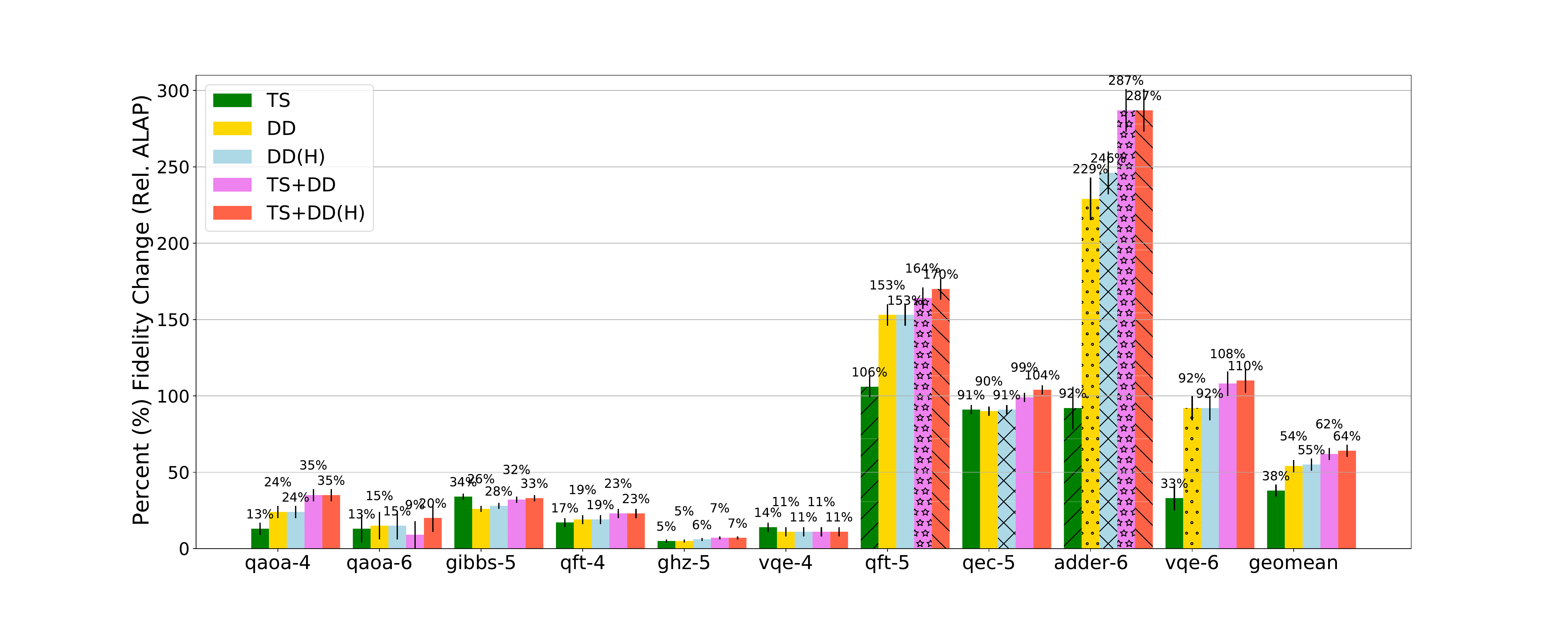}
\caption{Comparing POS benefits between TS, DD and TS+DD variants, relative to ALAP. While DD provides greater benefits than base TS, the error mitigation techniques interfere constructively and the combined approach, TS+DD, performs better than DD or DD(H). DD(H) has the highest average increase in POS. These results were generated with real QC experiments}
\label{fig:dd-experiment}
\end{figure}

Fig.\ref{fig:dd-experiment} shows a comparison of TS, DD, DD(H), TS+DD, and TS+DD(H).
Note that here, TS is abbreviated for conciseness and corresponds to the constrained TimeStitch approach, TS-SI+C, as it is practical for real-machine deployment.
These are detailed in Section \ref{5-m-evalcompare}.
All results are normalized to the ALAP baseline.

First, we note that although TS provides significant boosts in benchmark POS, the DD and DD(H) approaches perform equally or better than the constrained TS in all but two benchmarks.
The DD approaches are able to achieve geometric mean POS improvements of 54-55\% compared to the 38\% for constrained TS.
The primary reason is that DD can be employed in all slack windows while the constrained TS approach is limited to correcting windows which are allowed by practical circuit depth limitations (Section \ref{Proposed-C}). On the other hand, constrained DD does not require additional circuit instructions and provides error mitigation with operations already present within the original circuit.
It is worth noting from Fig.~\ref{Fig:Result-POS} that the unconstrained TS approach achieves a mean 50\% POS improvement which is more comparable with the DD benefits and is achieved without adding circuit gates.

While fidelity numbers indicate that DD can outperform TS, it is critical to note that the two approaches are not entirely mitigating the same errors. Additionally, their state refocusing methods differ, coming with their own trade-offs of either additional tuning or gate overhead. 
This means that rather than juxtaposing TS and DD as alternatives, there is most opportunity in employing the two techniques in conjunction to maximize the benefits of each.
Doing so provides considerable further improvements: TS+DD is able to achieve a 62\% POS improvement over ALAP on average, clearly highlighting the synergy between the two techniques.
Further, the use of TS and our heuristic threshold that selectively insert DD sequences based on window durations, TS+DD(H), pushes mean improvements to 64\% over the ALAP baseline.
Overall, the combined approach provides a 19\% mean relative improvement over the benefits of DD.

With the combined approach of TS+DD(H), POS improvements range from 7\% to a whopping 287\%, again dependent on circuit and machine characteristics.
These results clearly indicate that combining error mitigation techniques, even beyond those discussed here, can be a considerable fidelity thrust in the NISQ era, resulting in circuits optimized to their full potential.

\section{Discussion}
\label{7-discussion}

The TimeStitch framework targets slack, providing a solution for mitigating decoherence in quantum circuits without the need for additional gates in the final, optimized circuit. Additionally, TimeStitch presents the novel contribution of a slice+inverse tuning mechanism that respects QC frontiers and is enabled by quantum reversibility.

\subsection{Future Applications of TimeStitch}

Ensuring that quantum optimizations scale along with applications is critical. As discussed in Section~\ref{Infra}, current TimeStitch slack tuning overheads are manageable as they are contained within a single additional job that must be run before the execution of the final, rescheduled circuit. In near-term QCs where variational noise easily corrupts computation, this small overhead of slice+inverse tuning is trivial because the average fidelity improvements of +38\% on real QCs is significant, outweighing the additional job cost. In some cases, borderline POS values are brought well-above thresholds required for a definite solution because of TimeStitch. 

As devices scale, so will applications. As mentioned in Section~\ref{increased-idle-time-compiled-circuits}, the length of slack in compiled circuits will grow as QC executables increase in width and depth. In this study, circuits were modest of modest size, enabling thorough slack tuning on all windows. Overheads associated with slack tuning will be kept reasonable by TimeStitch in the future by carefully selecting critical or large windows for tuning. Additionally, the depth criteria can be strictly enforced to minimize the set of slice+inverse circuits included during optimization. Finally, this work searched for optimum slack schedules though an exhaustive search with a step size dependent on the number of experiments that can fit within a job, but future work will explore refined searching algorithms with lower sampling rates. Developing simulator models of these newly explored machine characteristics discovered by the novel SI tuning methods could also accelerate some of these research directions. 

\subsection{Related Quantum Proposals}

Past work includes the development of methodologies that impact decoherence in quantum circuits by reducing depth and thus overall circuit runtime~\cite{zulehner2019compiling,venturelli2018compiling,siraichi2018qubit,metodi2006scheduling,maslov2008quantum,childs2019circuit}. 
These works, however, although targeted to real QC topologies, do not consider variational QC characteristics such as gate error rates and gate durations for their techniques. 
There also exist frameworks that aim to decrease quantum circuit noise by taking device calibration data into consideration to improve program success~\cite{Tannu:2019a, vuillot2017error,murali2019noise}, but these techniques do not implement optimizations that take advantage of slack time in circuits. Next, optimizing schedulers exist that mitigate noise associated with crosstalk by considering device properties~\cite{ding2020systematic,murali2020software}. In addition, the methods of~\cite{zhang2020slackq} take advantage of quantum circuit slack, but focuses on the qubit mapping problem rather than error reduction on real QCs. 
Further, the benefits of our method complement other indirect decoherence mitigation approaches~\cite{li2019tackling,murali2019noise}. Finally, related work exists for both quantum reversibility and DD applied in quantum circuit optimization. These proposals and their relation to the TimeStitch framework are discussed in detail in Sections~\ref{Reversibility-applied} and~\ref{DD-considerations}, respectively.

\subsection{Exploiting Slack in Classical Computing}

At the circuit level, slack in a clock cycle can occur in the presence of conservative timing guardbands.
These have been exploited with multiple better-than-worse-case approaches~\cite{Ernst:2003,Gupta:2009,Tiwari:2007,Ravi:2018,Ravi:2019,Ravi:2020}.
Similarly, at the micro-architecture level, periods of time with less or no-activity can help save power at no additional performance costs.
These are often exploited via power/clock gating, multi threading~\cite{SMT}, instruction rescheduling~\cite{Fields:2002} and so on.

\section{Conclusion}
\label{8-conclusion}

Reducing the impact of decoherence is critical for substantial advancements on near-term QCs.
The unintentional coupling of qubits to their environment, and each other, adds significant noise to  computation,  and  improved  methods  to  combat  decoherence are  required  to  boost  the  performance  of  quantum  algorithms on real machines. This  article  presents  a  novel  technique that takes advantage of a largely unexplored space of quantum circuit slack, opening up a new domain of exploration. 

Quantum circuit slack will only become more prevalent in time.
Here, slack tuning improves the fidelity of compiled quantum circuits  without  either  increasing  total  gate  count  or introducing circuit partitioning that increases circuit duration.
By exploiting quantum reversibility and by constraining tuning circuits to the depth of the original application, we propose a practical design suited for a variety of applications and quantum machines, especially applications of low fidelity which are critical to improve. We evaluated our proposal TimeStitch on real quantum machines, on benchmarks that are critical to real-world quantum usecases. We additionally offer insights on challenges and optimizations suited to realistic deployment.

\bibliographystyle{ACM-Reference-Format}
\bibliography{refs}

\end{document}